# Electronic Structure and Carrier Mobilities of Twisted Graphene Helix.

Rajesh Thakur*[$], P. K. Ahluwalia[$], Ashok Kumar[&], B. Mohan[@], and Raman Sharma[$]

[$]*Department of Physics, Himachal Pradesh University, Summer Hill, Shimla 171005 Himachal Pradesh (INDIA)*

[&]Department of Physical Sciences, School of Basic and Applied Sciences, Central University of Punjab, Bathinda, 151001 Punjab (INDIA)

[@]Department of Physics, Govt. Degree College Sanjauli, Shimla-171006, *Himachal Pradesh (INDIA)*

(December, 2019)

* Corresponding author: Email: rajeshhputhakur@gmail.com

## Abstract

We have investigated the effect of twisting on electronic band structure, effective mass and carrier mobilities of three prototypes of AGNRs (N=6, 7 & 8) using Density functional theory combined with Deformation potential theory and Effective mass approximation. It is shown that the influence of twisted modes electron-phonon interaction is smaller than stretching modes, nonetheless, twisting has a profound effect on effective mass and mobilities. Similar to earlier reported conclusion in which the ideal N=3n wide HAGNR potentially exhibit an electron mobility comparable to intrinsic graphene, we also found that the ideal N=3n+2 HAGNRs hole mobility reside more closely to intrinsic graphene which could be increased further through Fluorine passivation. Thus, the control of the ribbon width along with passivation and extent of deformation are of paramount importance for determining the n-type or p-type of ribbons. Also, because of strong response to torsional strain, the N=8 F passivated AGNRs are the most appropriate for mechanical and high frequency switching. Our results suggest that twisting a ribbon can be considered as a good alternative way for controlled manipulation of the band structure and carrier mobilities for applications in mechanical switching devices.

# 1. Introduction

Development of the Graphene nanoribbons requires candidate systems with both a high carrier mobility and a sufficiently large electronic band gap which actually has been offered by the armchair graphene nanoribbons (AGNRs)[1–4]. The determination of electronic band structure and effect of uniform structural deformation is important to understand the electron transport of AGNRs either in pristine or strained form like twisted helical form. Twisting provides a novel way to tune band-gap and carrier mobilities and need to investigate thoroughly.

Although the GGA-PBE calculated band gap of GNRs [1] is underestimated as compared to measured band gap [5,6], yet the determination of other properties of AGNRs like effective mass[7,8] or work function[9] is observed to be in excellent agreement with experimental study[5,10]. Carrier's effective mass and mobility of Phosphorene nanoribbons (PNRs) has been determined with reasonable accuracy at GGA-PBE level of theory[11]. Because of the periodic oscillations of the band gap as a function of the magnitude of uniaxial strain in case of AGNRs[12], the simple application of Deformation Potential (DP) is not straightforward and the relative displacements of the bands energy surfaces in case of 1D system may be produced by twisting as studied in the case of carbon nanotube[13]. Therefore, it is interesting to investigate the extension of the DP equation for the band gap as a function of strain $E(\varepsilon_{ij}) = E_0 - E_1\Delta$, where $\Delta$ is the dilation $\Delta = \sum \varepsilon_{ij}$ and $E_1$ & $E_0$ is the DP & the band gap of unperturbed AGNRs respectively[14].

Not only the effect of twisting but also the effect of passivation on the electronic bands structure and effective mass of a twisted armchair of a bipartite lattice GNRs[15–18] have been studied within the fixed boundary condition making translational symmetry tractable only for a few discrete twist angles values ($\theta$). The systematic study of the narrow GNRs prototypes for **N**= 6, 7 & 8 each separated into basically three groups with a hierarchy of band gap response to torsional strain ($\theta$) given by $q = mod(N, 3)$ [19], $N$ is number of dimer lines across the ribbons width ($W$) has been taken out. We considered most frequently studied H-terminated AGNRs i.e HAGNRs and F-terminated AGNRs i.e FAGNRs[20–22] on the basis of their respective inversion and mirror symmetries[23]. Note that zigzag' type termination with N=odd have both mirror and inversion symmetry whereas zigzag termination with N=even have only mirror symmetry.

Futhermore, Al-Aqtash et. al. shown that the zigzag GNRs in their ground state is insensitive to twisting and no band gap change occurs[24] and substantial changes has been reported only for the narrowest AGNRs ribbons with W<1 nm[19]. In this paper, we report the carrier mobility of narrow twisted hydrogen and fluorine passivated AGNRs in this work. Absence of axial ionic relaxation allows us to study the helical with axial effective strain resulting from twisting as well as torsional strain simultaneously.

## 2. Computational Details

All the calculations in this work are performed using spin-polarized first-principles method by using SIESTA[25] package. Norm conserving Troullier Martin pseudo potential in fully separable Kleinman and Bylander form has been used to treat the electron-ion interactions[26]. The exchange and correlation energies have been treated within GGA-PBE functional[27]. The Kohn Sham orbitals were expanded as a linear combination of numerical pseudo atomic orbitals (LCAO) using a split-valence double-ζ basis set with double polarization functions (DZP). The convergence tolerance for total energy is chosen as $10^{-5} eV$ between two consecutive steps. Throughout geometry optimization, the confinement energy of numerical pseudo-atomic orbitals is taken as 0.01Ry. Minimization of energy was carried out using standard conjugate-gradient (CG) technique. Converged values of sampling for the k-mesh grid $\sim 10^{-2} Å^{-1}$ have been used according to Monkhorst-Pack scheme[28] to sample Brillioun zone.

The structures with fix lattice constant were relaxed until the force on each atom was less than $10^{-2} \, eVÅ^{-1}$. The spacing of the real space grid used to calculate the Hartree exchange and correlation contribution of the total energy and Hamiltonian was 800Ry for untwisted AGNRs and the converged values in a range between 1300Ry to 1450Ry was taken for twisted AGNRs. Vacuum region between two periodic images along non-periodic direction of about ~12Å were used in calculations to prevent the superficial interactions.

### 3. Results and Discussion

The lattice parameters of studied nanoribbons (HAGNRs and FAGNRs) are given in Table 1. The mechanic of transition of planar AGNR to helical shaped conformation is shown in Figure 1(a-b). Further increase in torsional angle decreased the unit cell size $L_M$ while increasing the torsional strain is demonstrated in Figure (1c-d). Likewise we have modeled and studied these few discrete torsional angles compatible with translation symmetry along Z-direction for HAGNRs and FAGNRs. We classified these different helical conformations by three parameters shown in Figure 1: lattice constant ($L_M$), torsional angle($\theta$), width ($W$) of AGNRs as given in the Table 1. The subscript $M$ of lattice constant $L_M$ is the number of times unit cell get repeated or multiplied to have required twisted super cell.

### 3.1. Electronic properties

To have an insight into the electronic properties of H & F AGNRs, we have performed band structure calculations. Because electromechanical response of twisted helical can be understood in a unified way by using the notion of effective strain, therefore, we will discuss the observable response only to effective strain in the rest of the article. The effective strain $\varepsilon^{eff} = 1/2(\theta^2 \Sigma_n^2)$ i.e. effective tensile strain associated with the twist angle(θ) and $\Sigma_n$ is the distance of $n^{th}$ dimer from axis[29]. The calculations reveal that among the untwisted AGNRs, the narrowest ones 6 H **&** F AGNR have the highest band-gap of 1.14eV and 1.41eV respectively, effectively making it a semiconductor. On applying a torsional strain the band gap increased monotonously in case of N=6 HAGNRs up to 2.01eV for extreme torsional strain which is the highest band gap among all the cases we have taken. But in case of N=6 FAGNRs, band gap does not vary much (See *Figure 2*).

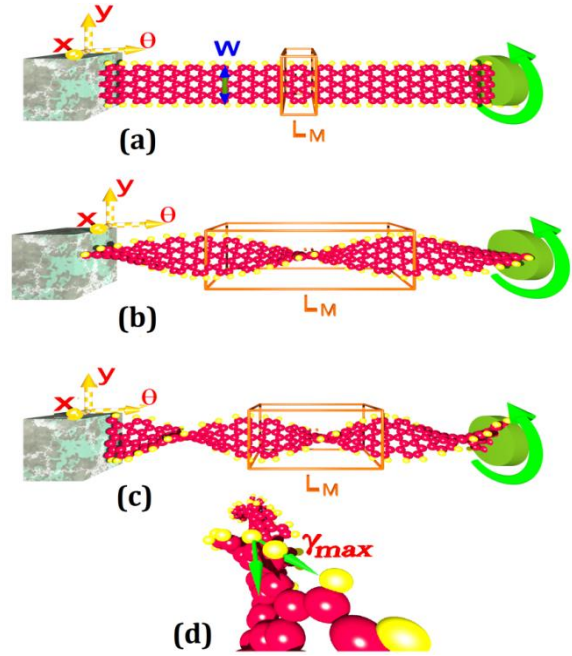

*Figure 1: Schematic description of mechanics of twisting that has been applied to the infinite long (a) planar GNR with minimum lattice constant $L_M$ and torsional angle θ=zero that morphs it into (b) helical shaped infinite long twisted AGNR with increased value of lattice constant and torsional angle that eventually shrinks to (c) the helical with much smaller lattice constant when gets more twisted. The $L_M$ is the lattice constant shown in cell, W is the width of AGNR and $\gamma_{max}$ is the angle between the rotation axis θ (pointing outward to this page) with the tangent to the curve joining the edge atoms.*

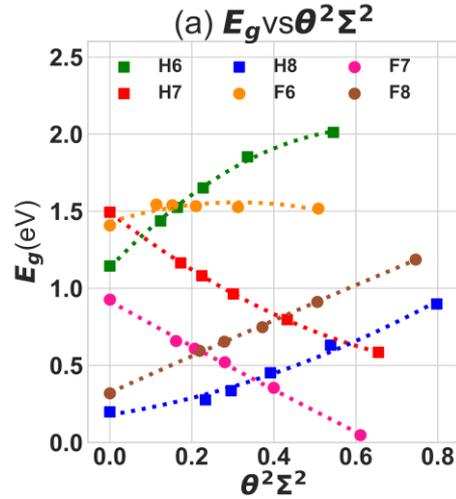

*Figure 2: Band gap as a function of effective strain for different width nanoribbons.*

The direct band gap symmetry point of untwisted N=6 HAGNRs although shifted from $\Gamma$ to $\pm K$. Likewise for N=8 H and F AGNRs helical conformations the response of the band gaps to effective strain vary monotonously as well as linearly with positive slope. Also, in these cases all the bands gaps are direct but interestingly no shift has been observed and band gap remain at $\Gamma$ point (Figure 5 and Figure 6). In effective strain space the trend of band gap response to torsional strain suggested the monotonous increasing behavior for **N**=6 & 8 HAGNRs (i.e. **q**=3N & **q**=3N+2), categorizing them into a one family. However, because of the effect of quantum size the band gap of **N**=6 AGNRs have larger gap values.

For N=7 H and F AGNRs case, the response of band gaps to the effective torsional strain as we have seen in N=8 case, is monotonous and linear however have negative slopes *Figure 2*. This kind of response categorize the **N**=7 AGNR (i.e. **q**=3N+1) into different family. The band gaps in case of untwisted 7-H and F AGNRs are direct at $\Gamma$ point (Figure 7a) and Figure 8a)), that shifted to $\pm K$ but remains direct even for higher torsional strain Figure 7(b to f) and Figure 8(b to f). In case of N=7 **&** 8 it is visibly obvious in *Figure 2* to see that the passivation with F shifts the band gap response to effective strain downward without altering the trend. The anomalous behavior of N=6 AGNRs can be understood through our previous study[34] in which we calculated the charges on H and F atoms **ΔQ**= 0.022e and **ΔQ** = -0.030e respectively that are not observed to vary on twisting and insignificantly vary (|0.002|) on varying dimer number $N$. Hence suggests that for narrower N=6 FAGNRs the electron density depletion from carbon skeleton is comparatively more.

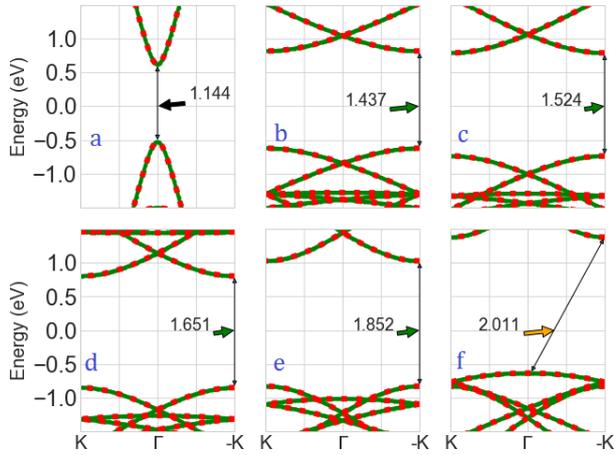
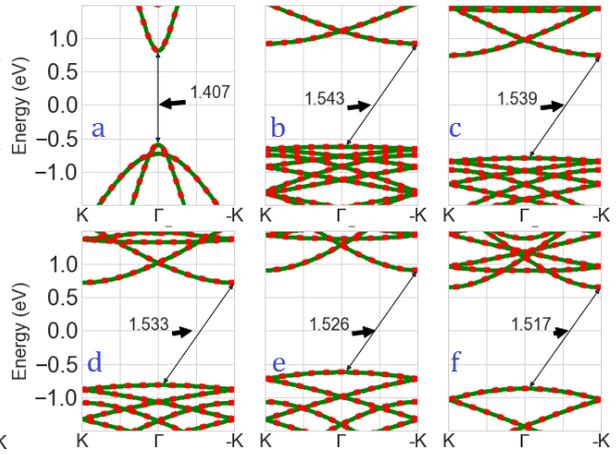

*Figure 3:* Calculated spin-polarized band structure of (a) Untwisted 6-HAGNR with zero torsional angle. Twisted 6-HAGNRs with torsional angle value (b) 0.097rad (c) 0.112rad (d) 0.132rad (e) 0.161rad (f) 0.207rad. The VBM and the CBM to measure the band gap value for each system is marked with arrows.

*Figure 4:* Calculated spin-polarized band structure of (a) Untwisted 6-FAGNR with zero torsional angle. Twisted 6-FAGNRs with torsional angle value (b) 0.094rad (c) 0.109rad (d) 0.128rad (e) 0.157rad (f) 0.202rad. The VBM and the CBM to measure the band gap value for each system is marked with arrows.

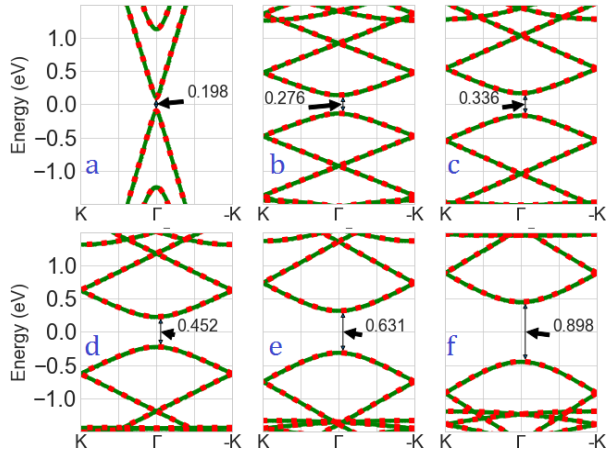
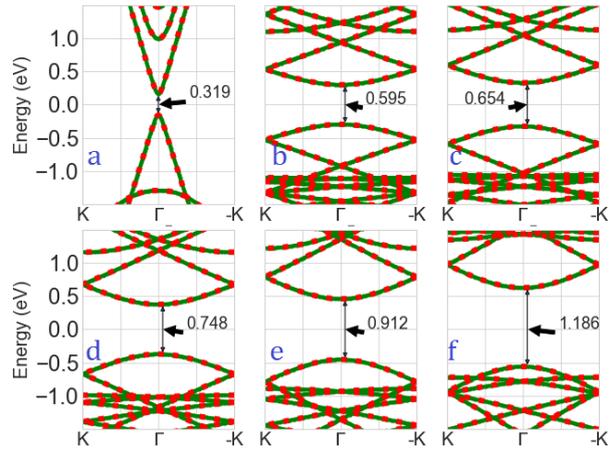

*Figure 5:* Calculated spin-polarized band structure of (a) Untwisted 8-HAGNR with zero torsional angle. Twisted 8-HAGNRs with torsional angle value (b) 0.086rad (c) 0.097rad (d) 0.112rad (e) 0.132rad (f) 0.162rad. The VBM and the CBM to measure the band gap value for each system is marked with arrows.

*Figure 6:* Calculated spin-polarized band structure of (a) Untwisted 8-FAGNR with zero torsional angle. Twisted 8-FAGNRs with torsional angle value (b) 0.084rad (c) 0.095rad (d) 0.110rad (e) 0.129rad (f) 0.158rad. The VBM and the CBM to measure the band gap value for each system is marked with arrows.

Furthermore, because of the downward shift of approximate value ~0.5eV in the band gap values of 7-FAGNRs w.r.t. 7-HAGNRs the Dirac cone has come to appear at $\pm K$ point.

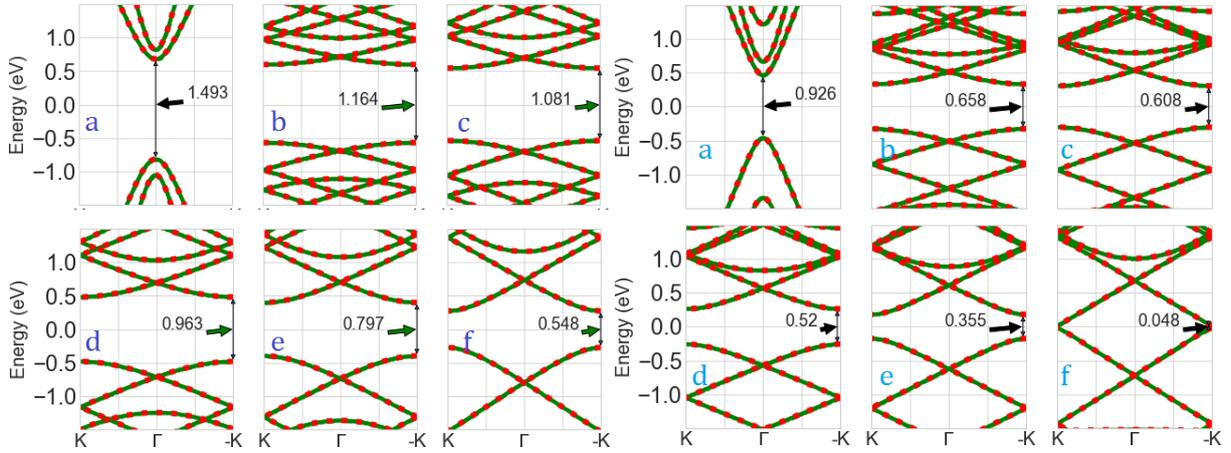

*Figure 7: Calculated spin-polarized band structure of (a) Untwisted 7-HAGNR with zero torsional angle. Twisted 7-HAGNRs with torsional angle value (b) 0.091rad (c) 0.104rad (d) 0.121rad (e) 0.146rad (f) 0.182rad. The VBM and the CBM to measure the band gap value for each system is marked with arrows.*

*Figure 8: Calculated spin-polarized band structure of (a) Untwisted 7-FAGNR with zero torsional angle. Twisted 7-FAGNRs with torsional angle value (b) 0.089rad (c) 0.101rad (d) 0.118rad (e) 0.142rad (f) 0.178rad. The VBM and the CBM to measure the band gap value for each system is marked with arrows.*

This allow the electrons with high momentum flow dissipation less through the strongly twisted 7=FAGNRs at low bias in contrast to the results of N=8 H and F AGNRs in which the untwisted nano-ribbons shows Dirac cone at $\Gamma$ point allowing the electrons with low momentum flows dissipation less through.

### 3.2. Effective Mass & Mobility

We calculated the carriers transport by applying deformation potential (DP) theory[14] and effective mass approximation. In earlier study of carbon nanotubes there is not only the stretching and breathing modes which contribute to the scattering but also shear or twist waves contributes to the scattering[13]. Similarly, in case of AGNRs the dilation term $\Delta=\sum \varepsilon_{ij}$ is the sum of longitudinal term and a correction, the twisting term. The longitudinal acoustic DP $E_{1L} = \delta(E_{edge} - eV_{Vaccum})/\delta\varepsilon_{xx}$, where $\delta E_{edge}$ is the shift of conduction band maxima energy (CBM) and valance band minima (VBM) is reference to the energy of vacuum level,

*Table 1:* **Structural parameters, Band gap ($E_g$), Effective masses at VBM ($|m_h^*|/m_e$) and CBM ($|m_e^*|/m_e$) and predicted carrier mobility for hole ($\mu_{1D}^h$) and electron ($\mu_{1D}^h$). Results of previous studies theoretical** [30][@], [31][#], [7][*#], [32][*] **and** [33][$]**; experimental** [10][&], [7][**]

| | $L$ Å | $\theta$ rad Å$^{-1}$ | $E_g$ eV | $|m_h^*|/m_e$ | $m_e^*/m_e$ | $\mu_{1D}^h$ cm$^2$V$^{-1}$s$^{-1}$ | $\mu_{1D}^e$ cm$^2$V$^{-1}$s$^{-1}$ | $W$ Å |
|---|---|---|---|---|---|---|---|---|
| **6-HAGNR** | 4.325$_{1.0}$ | Zero | 1.144 | 0.127, 0.12* | 0.125, 0.12* | 2.92x10$^2$ | 4.30x10$^4$ | 6.15, 6.15[#], 6.19[$] |
| | 32.439$_{7.5}$ | 0.097 | 1.437 | 0.133 | 0.133 | 2.74x10$^2$ | 3.93x10$^4$ | 6.133 |
| | 28.114$_{6.5}$ | 0.112 | 1.524 | 0.143 | 0.138 | 2.45x10$^2$ | 3.69x10$^4$ | 6.123 |
| | 23.789$_{5.5}$ | 0.132 | 1.651 | 0.166 | 0.159 | 1.96x10$^2$ | 3.0x10$^4$ | 6.110 |
| | 19.464$_{4.5}$ | 0.161 | 1.852 | 0.179 | 0.173 | 1.75x10$^2$ | 2.64x10$^4$ | 6.085 |
| | 15.139$_{3.5}$ | 0.207 | 2.011 | 0.303 | 0.237 | 7.91x10$^1$ | 1.65x10$^4$ | 6.030 |
| **7-HAGNR** | 4.32$_{1.0}$, 4.2& | Zero | 1.493 | 0.31, 0.33*[#], 0.41** | 0.35, 0.41*[#], 0.40** | 2.0x10$^3$ | 8.92x10$^1$ | 7.37, 7.31[#], |
| | 34.525$_{8.0}$ | 0.091 | 1.164 | 0.255 | 0.232 | 2.53x10$^3$ | 1.6x10$^2$ | 7.324 |
| | 30.209$_{7.0}$ | 0.104 | 1.081 | 0.2193 | 0.219 | 3.17x10$^3$ | 1.74x10$^2$ | 7.307 |
| | 25.894$_{6.0}$ | 0.121 | 0.963 | 0.2042 | 0.216 | 3.53x10$^3$ | 1.78x10$^2$ | 7.274 |
| | 21.578$_{5.0}$ | 0.146 | 0.797 | 0.1762 | 0.155 | 4.40x10$^3$ | 2.9 x10$^2$ | 7.224 |
| | 17.262$_{4.0}$ | 0.182 | 0.548 | 0.1332 | 0.151 | 6.70x10$^3$ | 3.05x10$^2$ | 7.131 |
| **8-HAGNR** | 4.31$_{1.0}$, 4.35[@] | Zero | 0.198 | 0.031,0.04* | 0.031, 0.04* | 4.0x10$^4$ | 2.42x10$^3$ | 8.63, 8.67[#], 8.66[$] |
| | 36.720$_{8.5}$ | 0.086 | 0.276 | 0.037 | 0.037 | 3.65x10$^4$ | 2.17x10$^3$ | 8.574 |
| | 32.325$_{7.5}$ | 0.097 | 0.336 | 0.046 | 0.046 | 2.64x10$^4$ | 1.57x10$^3$ | 8.564 |
| | 28.015$_{6.5}$ | 0.112 | 0.452 | 0.061 | 0.064 | 1.72x10$^4$ | 9.65x10$^2$ | 8.535 |
| | 23.705$_{5.5}$ | 0.132 | 0.631 | 0.089 | 0.097 | 9.82x10$^3$ | 5.14x10$^2$ | 8.493 |
| | 19.395$_{4.5}$ | 0.162 | 0.898 | 0.139 | 0.148 | 5.04x10$^3$ | 2.74x10$^2$ | 8.420 |
| **6-FAGNR** | 4.443$_{1.0}$ | Zero | 1.407 | 0.137 | 0.125 | 4.17x10$^2$ | 9.50x10$^3$ | 6.068 |
| | 33.322$_{7.5}$ | 0.094 | 1.543 | 0.781 | 0.260 | 30.6x10$^1$ | 3.17x10$^3$ | 6.059 |
| | 28.879$_{6.5}$ | 0.109 | 1.539 | 1.040 | 0.173 | 2.0x10$^1$ | 5.85x10$^3$ | 6.052 |
| | 24.436$_{5.5}$ | 0.128 | 1.533 | 1.017 | 0.161 | 2.1x10$^1$ | 6.49x10$^3$ | 6.046 |
| | 19.993$_{4.5}$ | 0.157 | 1.526 | 0.893 | 0.241 | 2.5x10$^2$ | 3.55x10$^3$ | 6.020 |
| | 15.551$_{3.5}$ | 0.202 | 1.517 | 0.837 | 0.299 | 2.76x10$^2$ | 2.58x10$^3$ | 5.967 |
| **7-FAGNR** | 4.421$_{1.0}$ | Zero | 0.926 | 0.179 | 0.176 | 8.67x10$^3$ | 2.98x10$^2$ | 7.263 |
| | 35.368$_{8.0}$ | 0.089 | 0.658 | 0.115 | 0.115 | 1.67x10$^4$ | 5.63x10$^2$ | 7.239 |
| | 30.947$_{7.0}$ | 0.101 | 0.608 | 0.101 | 0.101 | 2.04x10$^4$ | 6.87x10$^2$ | 7.222 |
| | 26.526$_{6.0}$ | 0.118 | 0.520 | 0.103 | 0.103 | 1.99x10$^4$ | 6.69x10$^2$ | 7.194 |
| | 22.105$_{5.0}$ | 0.142 | 0.355 | 0.074 | 0.074 | 3.27x10$^4$ | 1.10x10$^3$ | 7.141 |
| | 17.684$_{4.0}$ | 0.178 | 0.048 | massless | massless | -- | -- | 7.043 |
| **8-FHGNR** | $4.411_{1.0}$ | Zero | 0.319 | 0.046 | 0.046 | $1.82 \times 10^3$ | $4.15 \times 10^5$ | 8.540 |
| | $37.494_{8.5}$ | 0.084 | 0.595 | 0.080 | 0.076 | $8.0 \times 10^2$ | $1.94 \times 10^5$ | 8.510 |
| | $33.083_{7.5}$ | 0.095 | 0.654 | 0.094 | 0.089 | $6.27 \times 10^2$ | $1.51 \times 10^5$ | 8.494 |
| | $28.672_{6.5}$ | 0.110 | 0.748 | 0.110 | 0.110 | $4.94 \times 10^2$ | $1.11 \times 10^5$ | 8.469 |
| | $24.261_{5.5}$ | 0.129 | 0.912 | 0.139 | 0.139 | $3.47 \times 10^2$ | $7.77 \times 10^4$ | 8.424 |
| | $19.850_{4.5}$ | 0.158 | 1.186 | 0.188 | 0.208 | $2.22 \times 10^2$ | $4.26 \times 10^4$ | 8.351 |

$(eV_{Vaccum})$ is calculated for the considered AGNRs. The DP term for twisting mode has been obtained as $E_{1T} = \delta(E_{Edge} - eV_{Vaccum})/\delta\varepsilon^{eff}$ where $dE_{Edge}$ is the VBM for holes or the CBM for electrons. Within the elastic limits the magnitude of DP describes the degree interaction between electrons and phonons. Therefore, lower value of DP indicates a weaker electron phonon hence increase in the mobility of electrons (holes). Accordingly, for 1D systems the carriers' mobility is $\mu_{1D} = ne\hbar^2 C_{1D}/(2\pi k_B T)^{\frac{1}{2}}|m^*|^{3/2}E_1^2$,[35] where T=300K, and $C_{1D}$ is the stretching modulus caused by uniaxial strain which has been calculated using the expression $C_{1D} = \frac{1}{L_0}\frac{d^2}{d\varepsilon^2}E_S$ in which $E_S$ is the strain energy of a unit cell. In case of parabolic band the effective mass is related to how the band energy $(E_{nk})$ varies with the wave vector $(k)$ by $m^{*-1} = \hbar^{-2}\partial^2 E_{nk}/\partial k^2$ captures much of the physics of carrier transport.

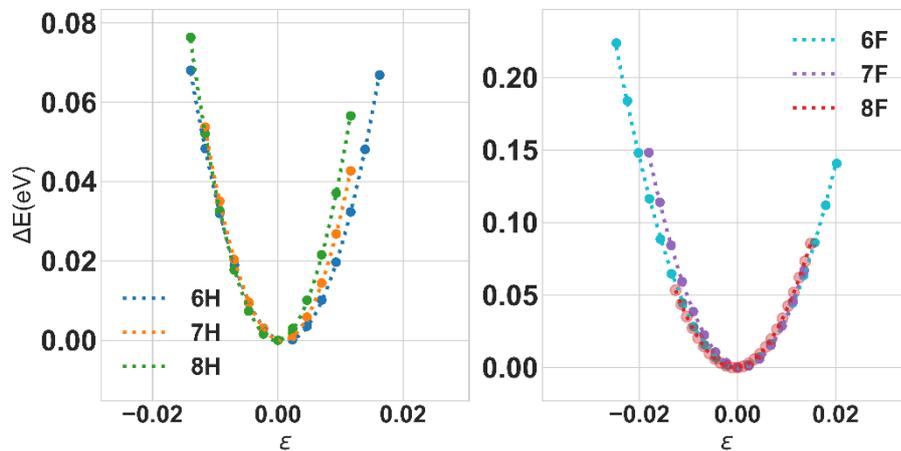

*Figure 9: Energy−strain relationship along armchair (a) HAGNR and (b) FAGNR.*

The variation of difference in strain energy (ΔE) with uniaxial strain (ε) applied along the armchair directions are shown in Figure 9. Based on those energy-strain curves we have obtained the line-stiffness $C_{1D}$ given in 2 which is in agreement with previous study[32]. The values of $C_{1D}$ are fairly equal for FAGNRs and HAGNRs counterparts. The magnitude of calculated E$_{1L}$ (2) for stretching mode for electrons and holes carries have been obtained by linear fitting of the response of band edges ($E_{Edge}$) i.e. valance band minima (VBM) and conduction band maxima (CBM) w.r.t. the vacuum potential to uniaxial strain. In earlier reported studies the value of E$_{1L}$ varies significantly (see ref.[32] & ref [2]) because of the different reference energy point taken in these studies. We followed the method described in ref [2] for determining the DP and the values which are in good agreement with the reported values.

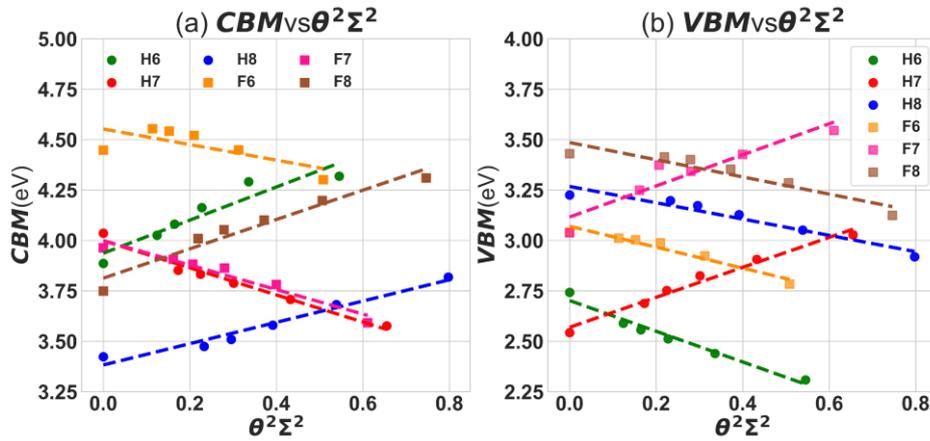

*Figure 10*: Shifts of valance band minima (VBM) and conduction band maxima (CBM) under effective torsional strain. The dashed lines are the linear fitting.

Figure 10 shows the shifts of band edges as a function of effective torsional strain along the armchair direction. Through twisting the AGNRs along the armchair directions, the DP constant $E_{IT}$ is then calculated as $dE_{Edge}/d\varepsilon^{eff}$, equivalent to the slope of the fitting lines, where $E_{Edge}$ is the energy of the conduction (valence) band edge. The $E_{IT}$ values of H and F AGNRs are shown in Table 2. Except N=6 FAGNRs, the standard deviation of all $E_{IT}$ values is smaller than 1%. The lattice scattering is determined by the shifts in the energy bands resulting from dilations associated with acoustics waves[14]. The higher value of shift occurred because of stretching mode, therefore the longitudinal DP values is taken for determining the mobilities, nonetheless, as we will see the effective mass response to twisting still has its profound effect on mobilities.

On the basis of obtained energy band spectrum and calculated values of $E_{1L}$, $E_{1T}$ and $C_{1D}$ we determined the acoustic phonon-limited mobility at room temperature (300 K). As already concluded that the scattering rate of carriers is dominated by in-plane stretching mode of AGNRs in comparison to twisted mode, nevertheless, in case of 6 HAGNR and 8 FAGNR the $E_{1L}$ for electron is 0.30 and -0.57, respectively. This implies that the electron-phonon interaction is very small which leads to very high mobility.

*Table 2: The line stiffness ($C_{1D}$), the longitudinal DP ($E_{1L}$) and torsional DP ($E_{1T}$) of Hydrogen and Fluorine passivated AGNRs.* [2]@ [32]#

|  | $C_{1D}$ ($10^{12}$ $eV/m$) | $E_{1L}$ of electron ($eV$) | $E_{1T}$ of electron ($eV$) | $E_{1L}$ of hole ($eV$) | $E_{1T}$ of hole ($eV$) |
|---|---|---|---|---|---|
| 6HAGNR | 1.38, 3.96# | 0.30, *0.84*@ | -0.76 | 9.11 | 0.82 |
| 7HAGNR | 1.66, 4.28# | 8.59, *9.36*@ | 0.74 | -2.01 | -0.67 |
| 8HAGNR | 1.88, 4.29# | 9.79, *10.41*@ | -0.40 | -2.39 | 0.53 |
| 6FAGNR | 1.61 | 1.74 | -0.52 | 7.76 | -0.39 |
| 7FAGNR | 1.88 | 8.23 | 0.77 | -1.51 | -0.61 |
| 8FAGNR | 1.66 | -0.57 | -0.43 | 8.53 | 0.73 |

Based on the obtained band structures, we calculate the effective mass of the charge carrier by parabolic fitting near the edge surface, which is presented in Table 1. It can be found that most of |m*|/m_e is small which means that the AGNRs have considerably high carrier mobility which is very sensitive to twisting. The responses of electron and holes mobilities are shown in the Figure 11. Our results show that the |m*|/m_e for electrons (holes) in untwisted 6, 7 & 8 HAGNRs are 0.13 (0.13), 0.34 (0.30) & 0.03 (0.03), respectively, which are in good agreement with Fischetti's report[2]. The value of $|m^*|/m_e$ for electrons and holes in untwisted

N=7 HAGNRs are also in good agreement with Senkovskiy et al[8] and Söde et al[7]. Furthermore, it is clearly seen from Table 1 that the effect of F passivation on the |m*|/m$_e$ of electron or hole for N= 7 & 8 is small, however, drastic change for N=6 is noticed. Our results also show the response of twisting for N=7, the $|m_e^*|/m_e$ ($|m_h^*|/m_e$) decreases from 0.34 to 0.15 (0.29 to 0.13) for HAGNRs and 0.18 to massless (0.18 to massless) for FAGNRs. But opposite trend is observed for N=6 & 7, the $|m_e^*|/m_e$ ($|m_h^*|/m_e$) values increased as a response to twisting shown in *Figure 11*. Furthermore, it is clearly seen that for even a small magnitude of twisting, the |*m*\*| of hole for N=6 FAGNRs becomes heavier than the smaller |*m*\*| of electron significantly, which means that the carrier transport ability of electron is very strong.

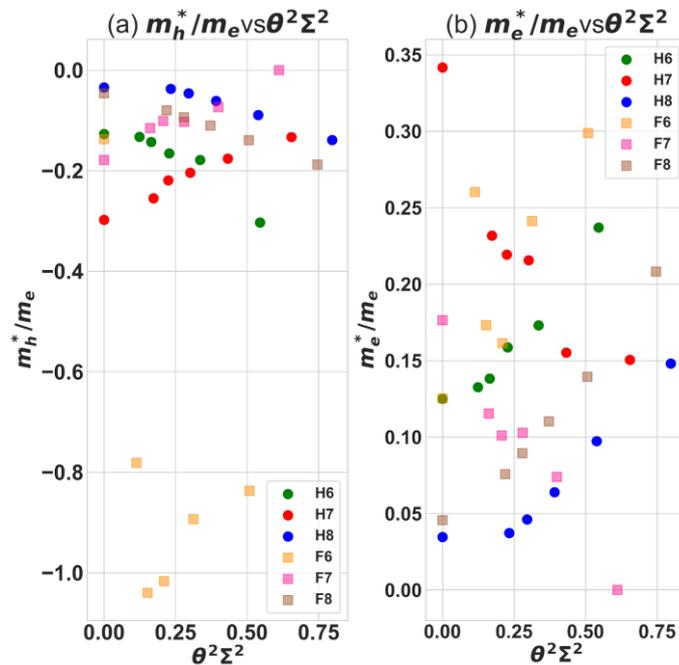

*Figure 11: The response of (a) holes and (b) electron effective mass to effective torsional strain.*

The electrons (holes) mobilities for N=6, 7 and 8 untwisted FAGNRs are about 9.5×10$^3$ (4.17×10$^2$), 2.98×10$^2$ (8.67×10$^3$) and 4.15×10$^5$ (1.82×10$^3$) cm$^2$V$^{-1}$s$^{-1}$, respectively. For N=6, 7 and 8 untwisted HAGNRs are about 4.3×10$^4$ (2.92×10$^2$), 8.92×10$^1$ (2.0×10$^3$) and 2.42×10$^3$ (4.0×10$^4$) cm$^2$V$^{-1}$s$^{-1}$, respectively are in good agreement with the electron mobilities of HAGNRs reported in ref [2]. Similar to earlier reported conclusion in which the ideal N=3n wide HAGNR were believed to potentially exhibit an electron mobility comparable to intrinsic graphene, we also found that the ideal N=3n+2 HAGNRs hole mobility is more close to intrinsic graphene. The mobility of electron (hole) could be increased further through F passivation up to $4.15\times10^5$ ($1.82\times10^3$) cm$^2$V$^{-1}$s$^{-1}$. Thus, the control of the ribbon width long with passivation are of paramount importance for determining the n-type or p-type of ribbons. Also N=8 FAGNRs has the best electron carrier transmitting capacity and the biggest difference between the electron and hole mobility among all considered cases that suggest its n-type characteristic.

For N=6 and 7 twisted AGNRs, the electron (holes) carriers mobility of FAGNRs are of the order ~0.1 (~0.1) times smaller and ~5 (~3) higher than its HAGNRs counterparts. Interestingly, for N=8 the electron (holes) carriers mobility of FAGNRs are of the order ~0.01 (~100) times than its HAGNRs counterparts. As a response to twisting the carriers mobilities decreased (Figure 12) for N= 6 and 8 for e.g. in HAGNRs, electrons (holes) from $4.30\times10^4$ ($2.92\times10^2$) to $1.65\times10^4$ ($7.91\times10^1$) and $2.42\times10^3$ ($4.0\times10^4$) to $2.74\times10^2$ ($5.04\times10^3$), whereas it get increased in case N=7 HAGNRs from $8.92\times10^1$ ($2.0\times10^3$) to $3.05\times10^2$ ($6.70\times10^3$). The most twisted configuration of N=7 FAGNRs has the Dirac cone formation at extreme twist resulting in the ballistic conductance and calculation of its mobilities is beyond the limitations of DP theory. It is clear from the Figure 12 that electron (holes) mobility for N=8 FAGNRs is $4.15\times10^5$ ($1.82\times10^3$) is comparable to graphene mobilities and also has the most sensitive and monotonous response to torsional strain. We find that the twisted AGNRs with appreciable band gap have a relatively large high mobilities (e.g. for HAGNRs~$4.30\times10^4$cm$^2$ V$^{-1}$ s$^{-1}$ and for FAGNRs ~$4.15\times10^5$cm$^2$ V$^{-1}$ s$^{-1}$), thus 'on–off' ratio that still way higher than TMDs (e.g. MoS2 ~500cm$^2$ V$^{-1}$ s$^{-1}$ [36]) and Black Phosphorene (~$2.6\times10^4$cm$^2$ V$^{-1}$ s$^{-1}$[37,38]) monolayer.

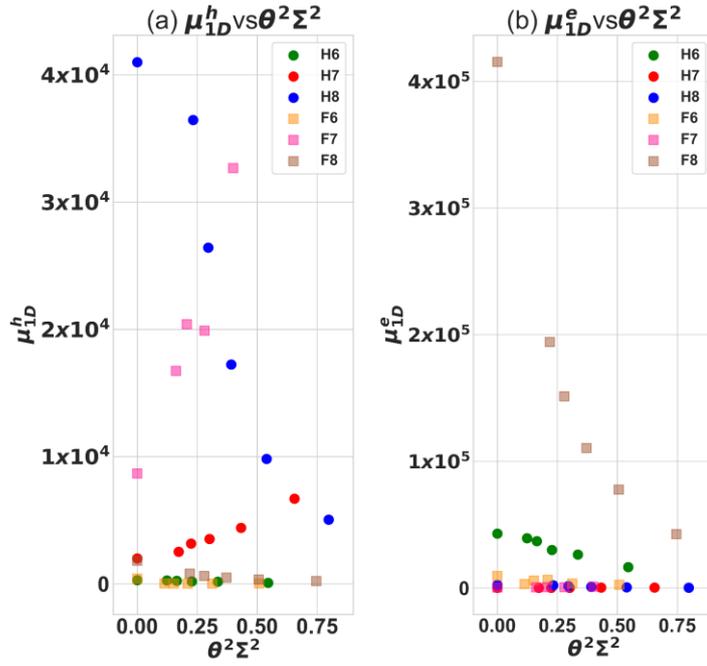

*Figure 12: The response of (a) holes mobilities and (b) electron mobilities to effective torsional strain.*

## 4. Conclusion

In summary, we have calculated the electronic band structures and the intrinsic charge carrier mobilities of H and F passivated AGNRs of three type of AGNRs, and the effect of twisting on them, using Density functional theory combined with DP theory and Effective mass approximation. We find that the directness of gap remain intact on twisting except in case of N=6 FAGNRs. The band varies monotonously as a response to torsional strain. The influences of twisted modes electron-phonon or hole-phonon interaction is smaller than stretching modes, nonetheless, twisting has a profound effect on effective mass and mobilities. The longitudinal DP are different for hole and electron for different AGNRs having order for CBM (VBM) as $E_{1L}^{3n+2} > E_{1L}^{3n+1} > E_{1L}^{3n}$ ($E_{1L}^{3n} > E_{1L}^{3n+2} > E_{1L}^{3n}$) for HAGNRs and as $E_{1L}^{3n+1} > E_{1L}^{3n} > E_{1L}^{3n+2}$ ($E_{1L}^{3n+2} > E_{1L}^{3n+1} > E_{1L}^{3n+1}$) for FAGNRs. The numerical results indicate that the electrons (holes) mobilities of N=6 and 7 twisted FAGNRs at room temperature, are of the order ~0.1 (~0.1) times smaller and ~5 (~3) higher than its HAGNRs counterparts. Interestingly, for N=8 the electron (holes) carriers mobility of FAGNRs are of the order ~0.01 smaller (~100 higher) times than its HAGNRs counterparts. The trend of electron (holes) mobility as $\mu_e^6 > \mu_e^8 > \mu_e^7$ ($\mu_h^8 > \mu_h^7 > \mu_h^6$) for HAGNRs and as $\mu_e^8 > \mu_e^6 > \mu_e^7$ ($\mu_h^7 > \mu_h^8 > \mu_h^6$) for FAGNRs. Due to the huge difference mobilities in hole and electron, N=8 & 6 FAGNRs can be considered as n-type semiconductors.

**Acknowledgements**

High performance computing facility of Centre for Development of Advanced Computing (CDAC), Pune and CVRAMAN, high performance computing cluster, at Himachal Pradesh University, Shimla has been used in obtaining the results presented in this paper. Author acknowledge the SIESTA TEAM for providing code under free license.